\documentclass[lettersize,journal]{IEEEtran}
\usepackage{amsmath,amsfonts}
\usepackage{algorithmic}
\usepackage{algorithm}
\usepackage{array}
\usepackage{textcomp}
\usepackage{stfloats}
\usepackage{url}
\usepackage{verbatim}
\usepackage{graphicx}
\usepackage{cite}
\usepackage{subfigure}
\usepackage{colortbl}
\usepackage{booktabs}
\usepackage{threeparttable}
\usepackage{pifont}
\usepackage{circledsteps}
\usepackage{tabularx}
\usepackage{multirow}
\usepackage{orcidlink}
\def\@@IEEEcomsocenforcemathfont{} 
\usepackage{mathptmx}

\hypersetup{hidelinks}
\begin{document}

\title{Tensor Manipulation Unit (TMU): Reconfigurable, Near-Memory Tensor Manipulation for High-Throughput AI SoC}

\author{
\IEEEauthorblockN{
Weiyu~Zhou\orcidlink{0009-0002-0035-3589}\IEEEauthorrefmark{2}\IEEEauthorrefmark{3},
\and
Zheng~Wang\orcidlink{0000-0003-2855-9570}\IEEEauthorrefmark{3}\IEEEauthorrefmark{1}\thanks{\IEEEauthorrefmark{1} Corresponding author: Zheng Wang, zheng.wang@siat.ac.cn},
\and
Chao~Chen\orcidlink{0000-0001-6488-224X}\IEEEauthorrefmark{3},
\and
Yike~Li\orcidlink{0009-0006-1140-1194}\IEEEauthorrefmark{3}\IEEEauthorrefmark{4},
\and
Yongkui~Yang\IEEEauthorrefmark{3},
\and
Zhuoyu~Wu\IEEEauthorrefmark{3},
\and
Anupam~Chattopadhyay\orcidlink{0000-0002-8818-6983}\IEEEauthorrefmark{6}
}
\\[0.8ex]
\IEEEauthorblockA{\IEEEauthorrefmark{2} Faculty of Science and Technology, University of Macau, Macau, China}
\\[0.8ex]
\IEEEauthorblockA{\IEEEauthorrefmark{3} Shenzhen Institutes of Advanced Technology, Chinese Academy of Sciences, Shenzhen, China}
\\[0.8ex]
\IEEEauthorblockA{\IEEEauthorrefmark{4} School of Electrical and Electronic Engineering, University College Dublin, Dublin, Ireland}
 \\[0.8ex]
\IEEEauthorblockA{\IEEEauthorrefmark{6} College of Computing and Data Science, Nanyang Technological University, Singapore}
}

\maketitle
\begin{abstract}
While recent advances in AI SoC design have focused heavily on accelerating tensor computation, the equally critical task of tensor manipulation—centered on high-volume data movement with minimal computation—remains underexplored. 
This work addresses that gap by introducing the Tensor Manipulation Unit (TMU): a reconfigurable, near-memory hardware block designed to execute data-movement-intensive (DMI) operators efficiently. TMU manipulates long datastreams in a memory-to-memory fashion using a RISC-inspired execution model and a unified addressing abstraction, enabling broad support for both coarse- and fine-grained tensor transformations. The proposed architecture integrates TMU alongside a TPU within a high-throughput AI SoC, leveraging double buffering and output forwarding to improve pipeline utilization. 
Fabricated in SMIC $40~\mathrm{nm}$ technology, the TMU occupies only $0.019~\mathrm{mm^2}$ while supporting over 10 representative TM operators. Benchmarking shows that TMU alone achieves up to $1413.43\times$ and $8.54\times$ operator-level latency reduction over ARM A72 and NVIDIA Jetson TX2, respectively. When integrated with the in-house TPU, the complete system achieves a $34.6\%$ reduction in end-to-end inference latency, demonstrating the effectiveness and scalability of reconfigurable tensor manipulation in modern AI SoCs.
\end{abstract}


\section{Introduction}
\IEEEPARstart{T}{he} 
previous decade has experienced the huge success of artificial neural networks (ANNs), which have been increasingly deployed on data centers, mobile edges, and terminal devices. The persistent efforts to improve peak performance and energy efficiency while reducing costs motivate the design and fabrication of customized chips for ANN. Systolic-Array-based Tensor Processing Unit (TPU) \cite{2017In} is specialized in executing integer operators during inference with massive parallelism, whereas General-purpose computing on Graphics Processing Unit (GPGPU) \cite{khairy2019survey} accelerates high-precision operators adopting vector-style floating-point units. Furthermore, any application inevitably contains highly flexible software code, which is conventionally deployed on a scalar-style RISC CPU. Consequently, a modern computing system or System-on-Chip (SoC) for artificial intelligence (AI) is ideally composed of scalar, vector, and tensor computing engines, plus the memory system and communication infrastructure \cite{liao2021ascend}.

Recently, the operators have become versatile in ANNs, and their types are far beyond general matrix multiplication (GEMM), convolution, and pooling. While most of the research studies focus on accelerating the compute-intensive operators \cite{chen2020survey}, very few studies and designs tend to address another rising performance bottleneck, caused by a series of data-move-intensive (DMI) operators. 
Existing mathematical frameworks give various names to such operators, where we name them tensor manipulation (TM), similar to array manipulation in \textit{NumPy} \cite{oliphant2006guide}. Detailed in Section \ref{sec:operators}, TM operators in general contain little to zero computation, but a large amount of data movement. They glue the compute-intensive operators by transforming tensors in ANN. 
NumPy contains 54 TM operators and the types are still growing. TM operators constitute key routines of math kernels, which are essential to pre-, post-, and intermediate processing of state-of-the-art ANN models.

\begin{figure}[t]
    \centering
    \includegraphics[width=1\linewidth]{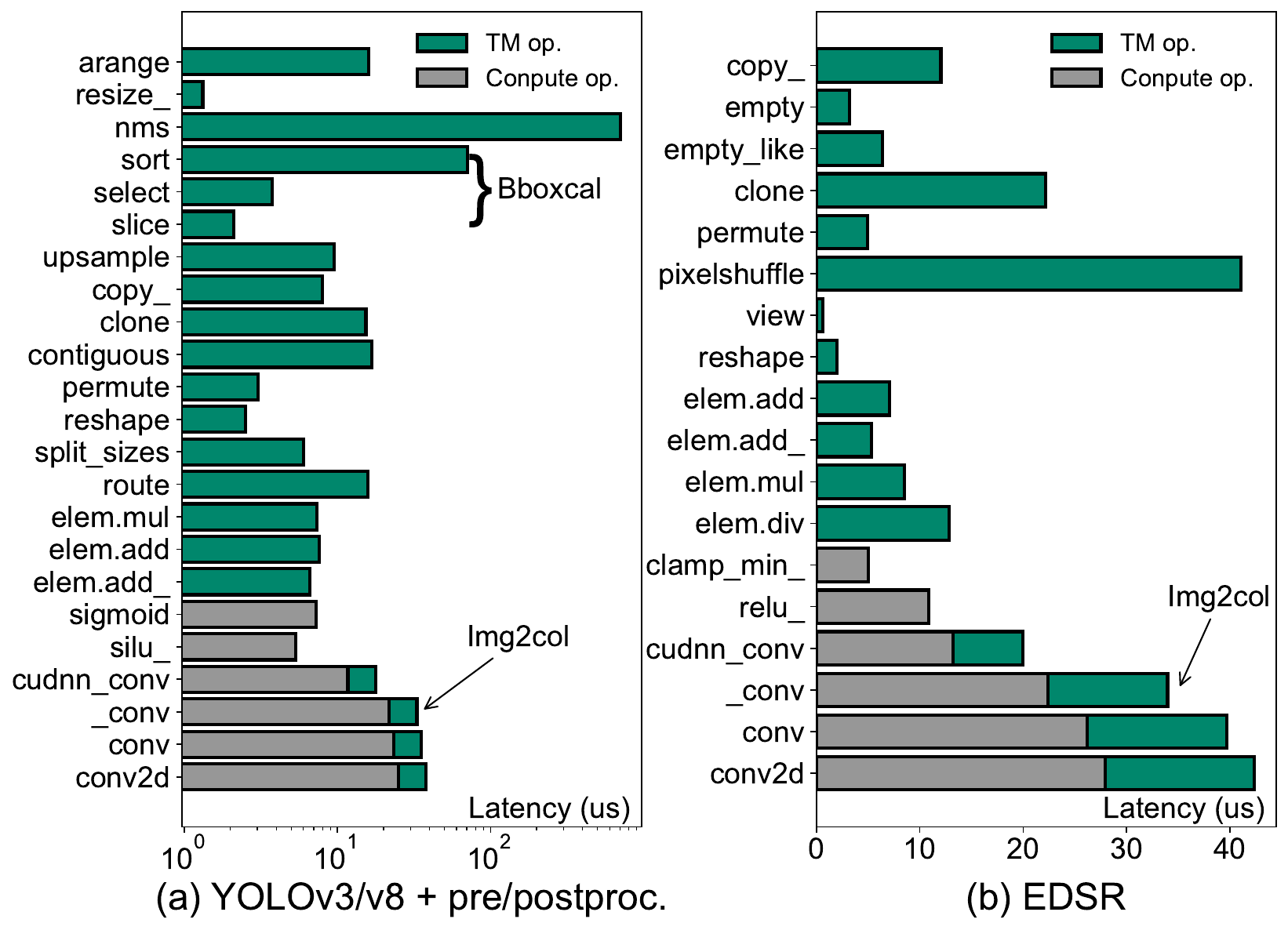}
    \caption{The average operator-level latency when running (a) YOLOv3/v8+pre/postproc and (b) EDSR on an NVIDIA RTX 3080.}
    \label{fig:operator profile}
\vspace{-0.5cm}
\end{figure}

As shown in Figure \ref{fig:operator profile}, we benchmark operator-level latency for object detection (YOLOv3/v8 \cite{redmon2018yolov3}\cite{varghese2024yolov8}) and super-resolution (EDSR \cite{lim2017enhanced}) NN models using an NVIDIA RTX 3080 GPU, which shows several TM operators are far more costly than compute operators. operators such as Bboxcal (realized by \textsl{sort}, \textsl{select} and \textsl{slice} in GPU) and non-maximum suppression (NMS) \cite{hsu2020ratio} constitute the post-processing phase of most YOLO-series NNs, where \textsl{pixelshuffle} is the key operator for the EDSR model. With regard to end-to-end inference latency, TM operators account for $40.62\%$ in EDSR when \textit{Img2col} is included, which is used in convolution and pooling for preparing activation buffers. The root cause of such long latency is that TM operators interact intensively with off-chip memory. However, most NN accelerators move data across layers of memory hierarchy to manipulate them inefficiently.

Effectively accelerating TM operators with minimal physical resources is of increasing importance, particularly as data movement has emerged as a widely recognized performance bottleneck in modern AI workloads. However, architectural support for improving TM performance remains limited. Academic efforts such as \cite{diniz2002data} propose FPGA-based data reorganization engines for inter-memory transfers, while \cite{zhang2019tucker} introduces a specialized accelerator for Tucker decomposition involving tensor permutations. Other approaches \cite{lloyd2015memory, akin2015data} explore in-memory data movement using 3D-stacked DRAM. Industry-wise, SoCs like NVIDIA’s Jetson and Rockchip’s RK3588 incorporate computer vision (CV) engines to manage tensor data, although their internal architecture details remain proprietary. Huawei’s Ascend \cite{liao2021ascend} includes a memory transfer engine (MTE), yet it only supports a limited subset of operators such as decomp, Img2col, and transpose.

A key limitation across prior designs is their reliance on fixed-function logic, which restricts adaptability to new and evolving TM operators. 
As TM grows increasingly diverse and complex, architectural reconfigurability becomes essential. Without a generalized and parameterized architectural abstraction, supporting these diverse patterns would require costly hardware redesigns or inefficient software fallbacks. 
This motivates the need for a reconfigurable TMU architecture that can flexibly adapt to heterogeneous TM behaviors while maintaining high throughput and low area overhead.

In this work, we propose \textbf{Tensor Manipulation Unit (TMU)} to support a series of TM operators on memory data streams. Other than a highly customized design, TMU is built under a \textbf{RISC-inspired execution model} which serves as a generic design template with reconfigurable registers to support a series of TM operators. A central feature of its design is the \textbf{Unified Address Abstraction}, which generalizes memory access patterns through parameterized address matrices. Thus operators can be encoded using shared fields with TM instructions.

TMU supports both coarse-grained and fine-grained data movements. In coarse-grained mode, the continuous datastream is uninterruptedly reorganized. In fine-grained mode, various modes of byte-level data assembling are achieved through a reconfigurable masking engine (RME). As an SoC component, TMU resides near the direct memory access (DMA) engine, which can either deliver manipulated tensor segments to tensor processing units (e.g., Img2col), or directly streaming back to off-chip DRAMs, therefore latency caused by data movement through memory hierarchies is reduced. Experiments on super-resolution and object detection NN models demonstrated a maximal speedup of $34.6\%$ in system inference latency when TMU couples with an in-house designed TPU \cite{li2024low} instead of ARM CPU coupling the same TPU. Individual TM operators can be executed orders of magnitude faster on TMU compared to embedded CPU and GPU. Furthermore, TMU occupies only $0.07\%$ silicon area of the in-house TPU, which proves its effectiveness and feasibility in modern SoCs.

The rest of the work is organized as follows. Section \ref{sec:related} surveys prior work on DMI operators. Section \ref{sec:operators} introduces typical TM operators. Section \ref{sec:method} details the design methodology for TMU. Section \ref{sec:architect} illustrates the system-level integration and microarchitectural details. Section \ref{sec:experiments} presents the benchmarking and synthesis results. Finally, Section \ref{sec:conclude} concludes the work.

\section{Related Work}\label{sec:related}
In neural networks, DMI operators—such as reshape, transpose, and slice—are essential for transforming tensors across layers, optimizing memory access, and enabling efficient parallelism. Although computationally lightweight, these operators often dominate runtime due to extensive memory traffic \cite{ivanov2021data}.

Early efforts leveraged FPGAs to construct data reorganization engines \cite{diniz2002data}, enabling coarse-grained data movement across memory hierarchies. Architectures like the Unstructured Data Processor (UDP) \cite{8778214} demonstrated improved memory efficiency through dynamic data transformation. Similarly, DMA profiling and optimization techniques on FPGAs \cite{8945776} reduced data transfer latency by overlapping movement with computation. Zhang et al. \cite{zhang2019tucker} proposed a dedicated accelerator for Tucker tensor decomposition; however, its domain specificity limits generalization to diverse DMI workloads.
IBM’s Active Messaging Engine (AME) \cite{sugawara2022data} is a lightweight, programmable DMA core integrated into the Power10 processor. While AMEs enable efficient offloading of data movement in high-performance computing (HPC) settings, their general-purpose messaging model lacks semantic support for tensor-centric DMI operators commonly found in deep learning.
ECNN \cite{huang2019ecnn} introduces a block-based accelerator optimized for energy-efficient CNN inference. It combines a hardware-oriented CNN model (ERNet) with a feature block instruction set (FBISA) to minimize external memory bandwidth. Although effective for fixed convolutional pipelines, its coarse execution granularity and lack of reconfigurable address control limit its applicability to dynamic TM operators such as Reshape and Bboxcal.

To reduce memory hierarchy overhead, in-situ data reorganization using 3D-stacked DRAM has been explored \cite{lloyd2015memory} \cite{akin2015data}. These solutions primarily target scientific and graph-based workloads and typically require custom memory stacks. TensorCIM \cite{tu202316} presents a digital computing-in-memory (CIM) architecture tailored for sparse tensor operations. Its design includes modules such as the redundancy-eliminated gathering manager (REGM) and input-lookahead CIM (ILA-CIM), significantly reducing off-chip and inter-chiplet traffic. Nevertheless, its specialization for sparse and irregular data patterns limits its utility for dense and fine-grained TM operators in standard neural networks.

Systolic arrays, widely used in CNN accelerators, encounter performance bottlenecks from memory bank conflicts introduced by irregular access patterns like those in Img2col. Recent approaches utilizing dynamic address generation \cite{10074771} improve memory utilization, though their benefits remain largely confined to convolution-heavy workloads and do not extend to the full range of TM operators.

Commercial AI SoCs, such as NVIDIA Jetson and Rockchip RK3588, incorporate proprietary vision engines that accelerate a limited set of TM operators. Similarly, Huawei’s Ascend AI processors \cite{liao2021ascend} feature an MTE capable of executing operators like Img2col and transpose, but lack extensibility to support emerging operators such as Bboxcal, PixelShuffle, and Route.

Despite these advancements, existing solutions suffer from three key limitations: (1) restricted operator coverage, (2) absence of reusable architectural abstractions, and (3) weak coupling to the memory subsystem, resulting in redundant data transfers and suboptimal bandwidth utilization.

To address these challenges, the proposed TMU provides a general-purpose, near-memory acceleration framework for DMI operators. TMU employs a RISC-inspired execution model with a programmable address generation engine, enabling efficient support for a wide spectrum of coarse- and fine-grained TM operators. Operating directly on memory data streams, TMU minimizes movement overhead. Integrated alongside a TPU within a heterogeneous SoC, TMU achieves up to $34.6\%$ system-level inference latency reduction while occupying only $0.07\%$ of TPU area, offering a scalable and reconfigurable solution for DMI optimization in modern AI workloads.

\begin{figure*}[th]
    \centering
    \includegraphics[width=1\linewidth]{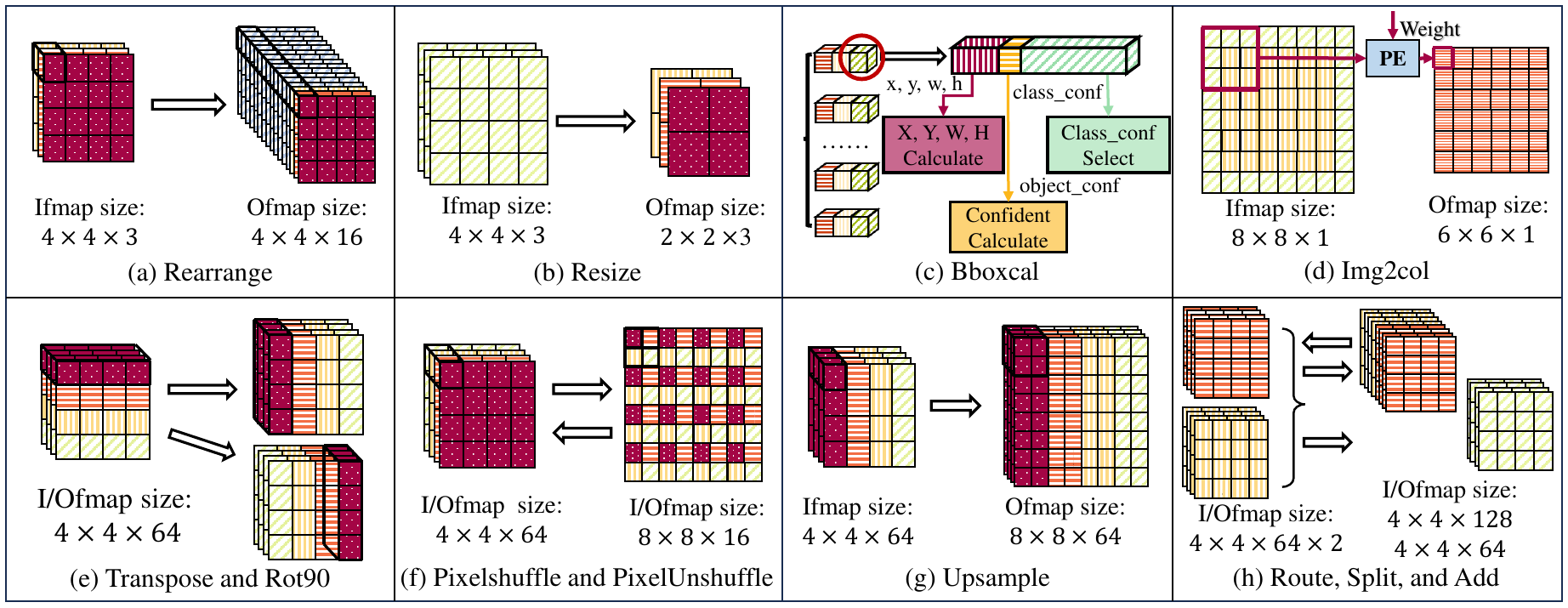}
    \caption{Graphical representation of typical TM operators adopted in state-of-the-art neural networks.}
    \label{fig:algorithm}
\end{figure*}

\section{Tensor Manipulation Operator}
\label{sec:operators}
TM operators are crucial for manipulating data structures in neural networks and scientific computing. Some TM operators involve computing routines, while others are purely data arrangements. The key TM operators that have been widely adopted in state-of-the-art neural networks and implemented in the proposed TMU are highlighted as follows.

\subsection{Fine-grained TM operator}
Fine-grained TM operators function at byte-level of granularity. They are crucial for optimizing memory access patterns and data layout in DNN pre-processing or post-processing, thereby improving throughput and efficiency. Key fine-grained operators include:
\subsubsection{Rearrange}
Shown in Fig. \ref{fig:algorithm}(a), Rearrange transforms RGB data streams into higher-channel feature maps (fmaps) (e.g., 16 channels) to favor AXI burst size and DRAM access patterns. It is crucial in the preprocessing stage of vision-based NN models, where byte-level data are \textit{fine-grained} rearranged.

\subsubsection{Resize}
As shown in Fig. \ref{fig:algorithm}(b), the bilinear interpolation operator calculates weighted averages of neighboring pixels to enable smooth image scaling. It is essential in vision AI models, ensuring visual quality with sub-pixel precision.

\subsubsection{Bboxcal}
As seen in Fig. \ref{fig:algorithm}(c), bounding boxes (Bboxes) with high confidence are extracted from YOLO's output tensors. Bboxcal is not typically accelerated by TPUs but significantly affects system inference latency. It operates on the byte level, making it a \textit{fine-grained} operator.

\subsubsection{Img2col}
Critical for speeding up variants of convolution and pooling, Img2col in Fig. \ref{fig:algorithm}(d) extracts necessary activations from input feature maps for the activation buffers in TPU.

\subsection{Coarse-grained TM operator}
Coarse-grained TM operators manage tensors or sub-tensors whose size exceeds the hardware-defined bus width (e.g., 16 bytes for a 128-bit AXI bus), emphasizing structural or dimensional transformations. They often reshape, reorient, or combine entire feature maps to meet the architectural requirements of DNN or to fuse information from different network paths. Important coarse-grained operators include:

\subsubsection{Transpose and Rot90}
Transpose operators rearrange the dimensions of tensors to adapt to specific tasks, while Rot90 rotates images by 90 degrees enhancing feature representation. These operators are widely used in DNNs for managing multidimensional data, as shown in Fig. \ref{fig:algorithm}(e) and (f).

\subsubsection{PixelShuffle and PixelUnshuffle}
PixelShuffle, shown in Fig. \ref{fig:algorithm}(g), rearranges feature maps for super-resolution, increasing width and height while reducing depth. Conversely, PixelUnshuffle, depicted in Fig. \ref{fig:algorithm}(h), reduces map dimensions as one of the downsampling operators.

\subsubsection{Upsample}
Illustrated in Fig. \ref{fig:algorithm}(i), upsample scales up feature maps, maintaining depth while expanding width and height, pivotal in object detection and segmentation.

\subsubsection{Route, Split, and Add}
Route (also known as Concat) combines feature maps along the channel dimension as depicted in Fig. \ref{fig:algorithm}(j). Split divides feature maps along the channel dimension as shown in Fig. \ref{fig:algorithm}(k). Add (also known as residual layer) performs element-wise additions with two tensors, which is essential for residual networks as depicted in Fig. \ref{fig:algorithm}(l).

\section{Tensor Manipulation Methodology} \label{sec:method}
To raise the abstraction level of the design, we have introduced a generic tensor manipulation methodology including the execution model and address generation. The growing number of TM operators can fit into the design template following the approach outlined in this section.

\begin{figure}
    \centering
    \includegraphics[width=1\linewidth]{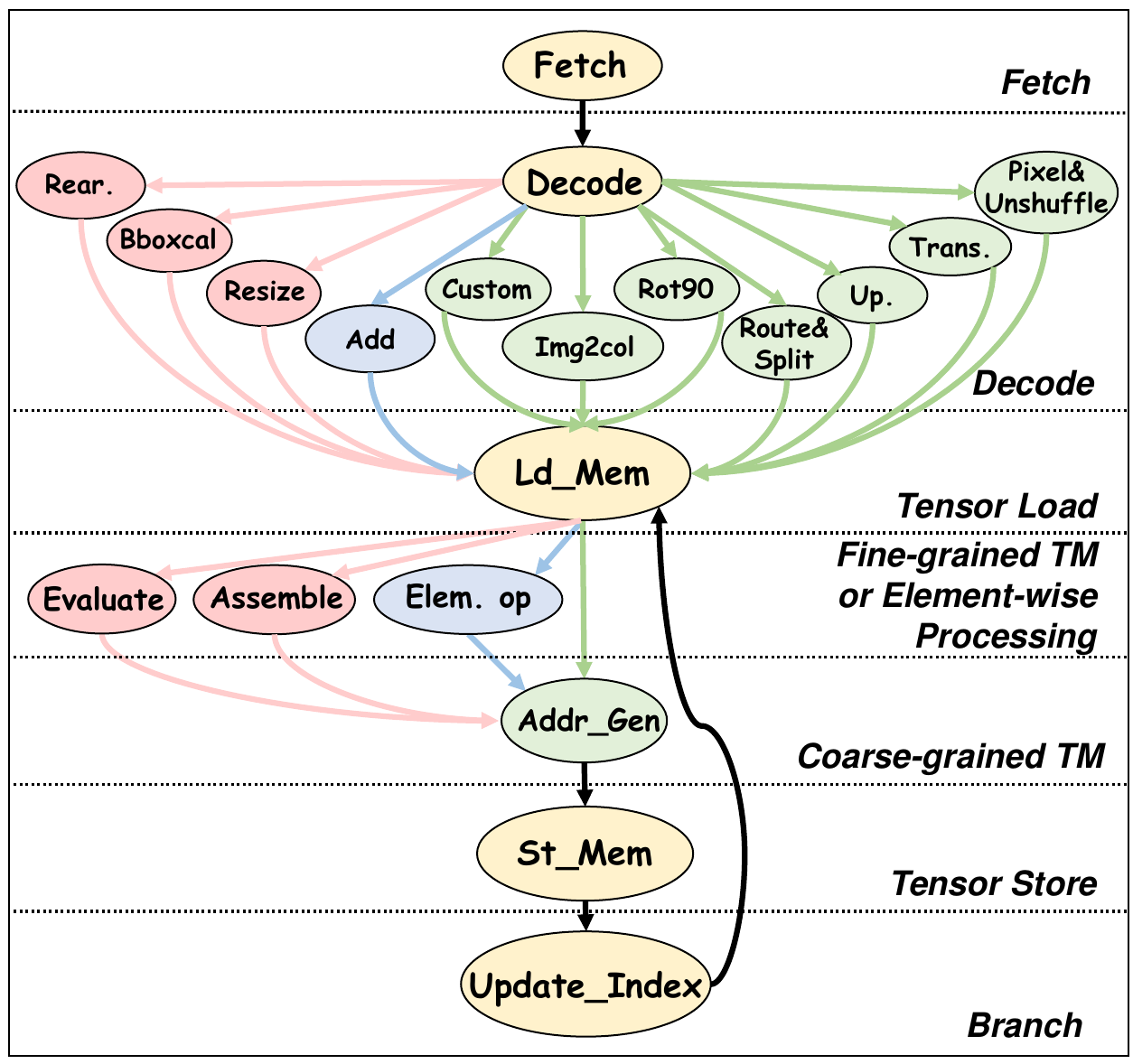}
    \caption{Generic execution model for TM}
    \label{fig:execution_model}
\vspace{-0.4cm}
\end{figure}

\subsection{Generic Execution Model}
We propose a RISC-inspired execution model for TMU that abstracts the tensor manipulation process into eight configurable stages, as illustrated in Fig. \ref{fig:execution_model}. Each stage represents a distinct class of dataflow transformation or control behavior and can be selectively activated based on the characteristics of a given TM operator. This stage-based abstraction allows diverse tensor manipulations to be expressed uniformly within a unified operational framework. The functionalities of each stage are described as follows.

\subsubsection{Fetch}
TMU acquires instructions from local storage.

\subsubsection{Decode}
The instruction is issued to determine its functionality.

\subsubsection{Tensor Load}
TMU initiates the loading of required tensors from memory, typically from off-chip DRAM.

\subsubsection{Fine-grained TM}
Certain TM operators—such as Resize, Rearrange, and Bboxcal—manipulate data at the \textit{byte-level}, often requiring flexible selection and reorganization of fine-grained tensor elements. These operations can be categorized into two high-level modes: \textit{assemble}, which gathers selected bytes and packs them into a continuous output stream; and \textit{evaluate}, which filters bytes based on simple comparison or thresholding and forwards only those of interest. 

\subsubsection{Element-wise Processing}
This category includes TM operators such as Add and Mul, which perform element-wise computations across corresponding tensor positions. Each tensor element is processed independently, making these operations highly parallelizable and well-suited for fusing with adjacent computation stages. 

\subsubsection{Coarse-grained TM}
The remaining TM operators, such as Transpose, PixelShuffle, and Split, operate on entire tensor blocks and perform structured layout transformations defined by tensor strides, shapes, and dimensions. Their access behavior can be uniformly described using a pattern-driven addressing abstraction, which encodes these transformations through a shared matrix-based formulation. This abstraction forms the foundation of the address generator discussed in the following section \ref{sec:addr_gen}.

\subsubsection{Tensor Store}
Manipulated tensors are written back to memory (e.g., off-chip DRAM) or forwarded to downstream computing engines (e.g., TPU) for subsequent processing.

\subsubsection{Branch}
Long tensors cannot be manipulated in a single run. This stage updates the address and loads the following tensor segment, before proceeding to the next instruction.

Taking advantage of the above TM execution model, designers can implement various types of TM operators and enhance the versatility of the TMU. The design template also facilitates resource sharing, where there is a large similarity between TM operators. It is important to note that although our model follows a RISC-inspired pipeline structure, it is not related to or compatible with standard RISC instruction sets. 

\subsection{Unified Address Abstraction for Tensor Manipulation}\label{sec:addr_gen}
A key feature of the proposed TMU is its ability to perform tensor manipulation directly in a memory-to-memory fashion, eliminating the need for CPU-driven address computations. To support a wide range of coarse-grained TM operators, the TMU employs a unified address abstraction that models memory access patterns through a parameterized and reusable formulation.

In this abstraction, each operator’s access pattern is represented as an affine transformation from input tensor indices to output memory locations. Rather than implementing dedicated address generation logic for each operator, the TMU employs a shared matrix-based approach that captures stride, padding, scaling, and layout reordering in a common structure.

\begin{table}[htbp]
\caption{Parameters used in Eq. 1 and Table \uppercase\expandafter{\romannumeral2}.}
\label{tab:param eq}
\begin{minipage}[t]{0.24\textwidth}
        \centering
        \begin{tabular}{cc}
        \toprule
        \textbf{Abbr.} & \textbf{Meaning}\\
        \midrule
        $addr_{out}$&Access address \\
        $addr_{base}$&Base address\\
        $x_{i}$&Ifmap X Position\\
        $y_{i}$&Ifmap Y Position\\
        $c_{i}$&Ifmap C Position\\
        $x_{o}$&Ofmap X Position\\
        $y_{o}$&Ofmap Y Position\\
        $c_{o}$&Ofmap C Position\\
        \bottomrule
        \end{tabular}
\end{minipage}
\begin{minipage}[t]{0.24\textwidth}
        \centering
        \begin{tabular}{cc}
        \toprule
        \textbf{Abbr.} & \textbf{Meaning}\\
        \midrule
        $w_{i}$&Ifmap Width\\
        $x_{k}$&Fmap Kernel Width\\
        $y_{k}$&Fmap Kernel Height\\
        $x_{p}$&Fmap Padding Width\\
        $y_{p}$&Fmap Padding Height\\
        $x_{s}$&Fmap Stride Width\\
        $y_{s}$&Fmap Stride Height\\
        $s$&Fmap Scale Factor\\
        \bottomrule
        \end{tabular}
\end{minipage}
\end{table}

The unified address abstraction dynamically computes source and destination addresses at runtime, guided by operator-specific parameters provided via the instruction stream. These parameters instantiate transformation matrices that formalize index mappings, as shown in Eq. \ref{eq:0}, with associated terms defined in Table \ref{tab:param eq}.

\begin{equation}
\begin{aligned}
addr_{out}=addr_{base}+y_o\times c_o+x_o\times c_o\\
\begin{aligned}
\begin{pmatrix}  
  x_{o} \\  
  y_{o} \\  
  c_{o} \\
\end{pmatrix} 
=
\begin{pmatrix}  
  a_{11} & a_{12} & a_{13} \\  
  a_{21} & a_{22} & a_{23} \\  
  a_{31} & a_{32} & a_{33}  
\end{pmatrix} 
\begin{pmatrix}  
  x_{i} \\  
  y_{i} \\  
  c_{i}  
\end{pmatrix} 
+\begin{pmatrix}  
  b_{1} \\  
  b_{2} \\  
  b_{3} \\
\end{pmatrix} 
\end{aligned}
\end{aligned}
\label{eq:0}
\end{equation}

Each TM operator corresponds to a specific pair of transformation matrices $(\mathbf{A}, \mathbf{B})$, which encode the linear relationship between input and output index triplets. For three-dimensional feature maps, the elements of $\mathbf{A}$ and $\mathbf{B}$ are typically constants, while the output indices $(x_o, y_o, c_o)$ are computed as functions of the input indices $(x_i, y_i, c_i)$, adapted to the semantics of each TM operator.
Table \ref{tab:matrix eq} lists representative configurations of $\mathbf{A}$ and $\mathbf{B}$ for commonly used coarse-grained operators, including Transpose, Img2col, and PixelUnshuffle. This matrix-based abstraction enables flexible support for strided and dilated access, channel fusion or splitting, and other compound tensor transformations.

By encoding transformation parameters directly into TMU instruction fields, this scheme allows runtime interpretation and execution without hardware modification. As a result, the same address generation datapath can support a wide range of tensor manipulation behaviors through lightweight reconfiguration.

\begin{table}[htbp]
    \centering
    \caption{The A and B matrices of address generation for coarse-grained TM operators.}
    \label{tab:matrix eq}
    \begin{tabular}{p{1.5cm}p{6.5cm}}
        \toprule
        \textbf{TM Op.} & \multirow{1}{=}{\centering\textbf{Eq.}}\\
        \midrule
        Transpose&
            $\begin{aligned}
            \begin{pmatrix}  
              x_{o} \\  
              y_{o} \\  
              c_{o} \\
            \end{pmatrix} 
            =
            \begin{pmatrix}  
              0 & 1 & 0 \\  
              w_i & 0 & 0 \\  
              0 & 0 & 1  \\
            \end{pmatrix} 
            \begin{pmatrix}  
              x_{i} \\  
              y_{i} \\  
              c_{i}  
            \end{pmatrix} 
            \end{aligned}$\\
        \addlinespace[1ex]
        Rot90&
            $\begin{aligned}
            \begin{pmatrix}  
              x_{o} \\  
              y_{o} \\  
              c_{o} \\
            \end{pmatrix} 
            =
            \begin{pmatrix}  
              0 & -1 & 0 \\  
              w_i & 0 & 0 \\  
              0 & 0 & 1  \\
            \end{pmatrix} 
            \begin{pmatrix}  
              x_{i} \\  
              y_{i} \\  
              c_{i}  
            \end{pmatrix} 
            +\begin{pmatrix}  
              w_i \\  
              0 \\  
              0 \\
            \end{pmatrix} 
            \end{aligned}$\\
        \addlinespace[1ex]
        Img2col&
        $\begin{aligned}
        \begin{pmatrix}  
          x_{o} \\  
          y_{o} \\  
          c_{o} \\
        \end{pmatrix} 
        &=
        \begin{pmatrix}  
          \frac{1}{x_{s}}  & 0 & 0 \\  
          0 & \frac{w_i}{y_{s}} & 0 \\  
          0 & 0 & 1  
        \end{pmatrix} 
        \begin{pmatrix}  
          x_{i} \\  
          y_{i} \\  
          c_{i}  
        \end{pmatrix} 
        +
        \begin{pmatrix}  
          \frac{2\times x_{p}-x_{k}}{x_{s}} + 1 \\  
          \frac{2\times y_{p}-y_{k}}{y_{s}} + 1 \\  
          0 \\
        \end{pmatrix} 
        \end{aligned}$\\
        \addlinespace[1ex]
        PixelShuffle&
        $\begin{aligned}
        \begin{pmatrix}  
          x_{o} \\  
          y_{o} \\  
          c_{o} \\
        \end{pmatrix} 
        =
        \begin{pmatrix}  
          1 & 0 & 0 \\  
          0 & s\times w_i & 0 \\  
          0 & 0 & \frac{1}{s}  
        \end{pmatrix} 
        \begin{pmatrix}  
          x_{i} \\  
          y_{i} \\  
          c_{i}  
        \end{pmatrix} 
        \end{aligned}$\\
        \addlinespace[1ex]
        PixelUnshuffle&
        $\begin{aligned}
        \begin{pmatrix}  
          x_{o} \\  
          y_{o} \\  
          c_{o} \\
        \end{pmatrix} 
        =
        \begin{pmatrix}  
          s & 0 & 0 \\  
          0 & w_i & 0 \\  
          0 & 0 & 1  
        \end{pmatrix} 
        \begin{pmatrix}  
          x_{i} \\  
          y_{i} \\  
          c_{i}  
        \end{pmatrix} 
        \end{aligned}$\\
        \addlinespace[1ex]
        Upsample&
        $\begin{aligned}
        \begin{pmatrix}  
          x_{o} \\  
          y_{o} \\  
          c_{o} \\
        \end{pmatrix} 
        =
        \begin{pmatrix}  
          s & 0 & 0 \\  
          0 & s\times s\times w_i & 0 \\  
          0 & 0 & 1  
        \end{pmatrix} 
        \begin{pmatrix}  
          x_{i} \\  
          y_{i} \\  
          c_{i}  
        \end{pmatrix} 
        \end{aligned}$\\
        \addlinespace[1ex]
        Route&
        $\begin{aligned}
        \begin{pmatrix}  
          x_{o} \\  
          y_{o} \\  
          c_{o} \\
        \end{pmatrix} 
        =
        \begin{pmatrix}  
          1 & 0 & 0 & 0 \\  
          0 & w_i & 0 & 0 \\  
          0 & 0 & 1 & 1 
        \end{pmatrix} 
        \begin{pmatrix}  
          x_{i} \\  
          y_{i} \\  
          c_{i1} \\
          c_{i2} 
        \end{pmatrix} 
        \end{aligned}$\\  
        \addlinespace[1ex]
        Split&
        $\begin{aligned}
        \begin{pmatrix}  
          x_{o} \\  
          y_{o} \\  
          c_{o} \\
        \end{pmatrix} 
        =
        \begin{pmatrix}  
          1 & 0 & 0 \\  
          0 & w_i & 0 \\  
          0 & 0 & \frac{1}{s} 
        \end{pmatrix} 
        \begin{pmatrix}  
          x_{i} \\  
          y_{i} \\  
          c_{i}
        \end{pmatrix} 
        \end{aligned}$\\
        \addlinespace[1ex]
        Add&
        $\begin{aligned}
        \begin{pmatrix}  
          x_{o} \\  
          y_{o} \\  
          c_{o} \\
        \end{pmatrix} 
        =
        \begin{pmatrix}  
          1 & 0 & 0 \\  
          0 & w_i & 0 \\  
          0 & 0 & 1 
        \end{pmatrix} 
        \begin{pmatrix}  
          x_{i} \\  
          y_{i} \\  
          c_{i}
        \end{pmatrix} 
        \end{aligned}$\\
        \addlinespace[1ex]       
        \bottomrule
    \end{tabular}
\vspace{-0.4cm}
\end{table}
\section{TMU Architecture} \label{sec:architect}
This section presents the implementation of the TMU, covering its system-level integration, microarchitecture, and dataflow organization.

\subsection{System Architecture}
\begin{figure}[!t]
    \centering
    \includegraphics[width=1\linewidth]{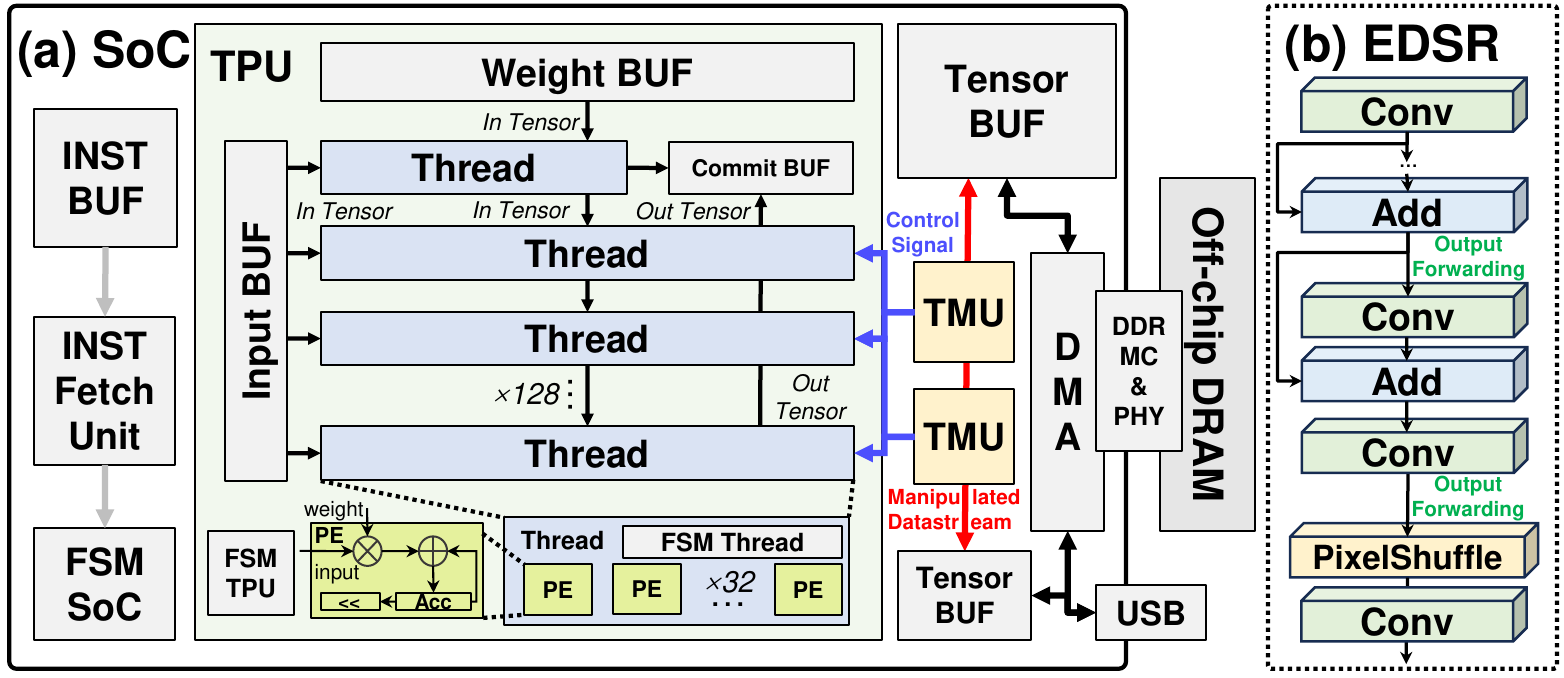}
    \caption{(a) System architecture integrating the proposed TMU and TPU. Two TMUs are deployed to support tensor prefetching. (b) Network architecture of EDSR.}
    \label{fig:architecture}
\end{figure}

As illustrated in Fig. \ref{fig:architecture}, the TMU serves as a key system-level component within our AI-oriented SoC architecture. The SoC executes neural networks as a sequence of operators, orchestrated via an instruction-driven execution model. It integrates both a TPU for compute-intensive tasks (e.g., conv, see Fig. \ref{fig:architecture}(b)) and a TMU for TM operators.
The TPU features a systolic array comprising 128 threads, each consisting of 32 PEs. Each PE is equipped with a 9-bit signed multiplier. Input tensors are fetched from DRAM through an AXI4 interconnect, buffered, and broadcast to the threads. The commit buffer aggregates the results from all threads and transfers them back to DRAM.
In contrast, the TMU is optimized for \textit{data-movement-intensive} and \textit{compute-light} operators. It reshapes, rearranges, or redirects datastreams retrieved from DRAM, and either writes them back to memory or forwards them to other computational threads. Forwarded datastreams can support element-wise operations (e.g., Add) or non-regular TM operators such as PixelShuffle (see Fig. \ref{fig:architecture}(b)).

\begin{figure}[!b]
    \centering
    \includegraphics[width=1\linewidth]{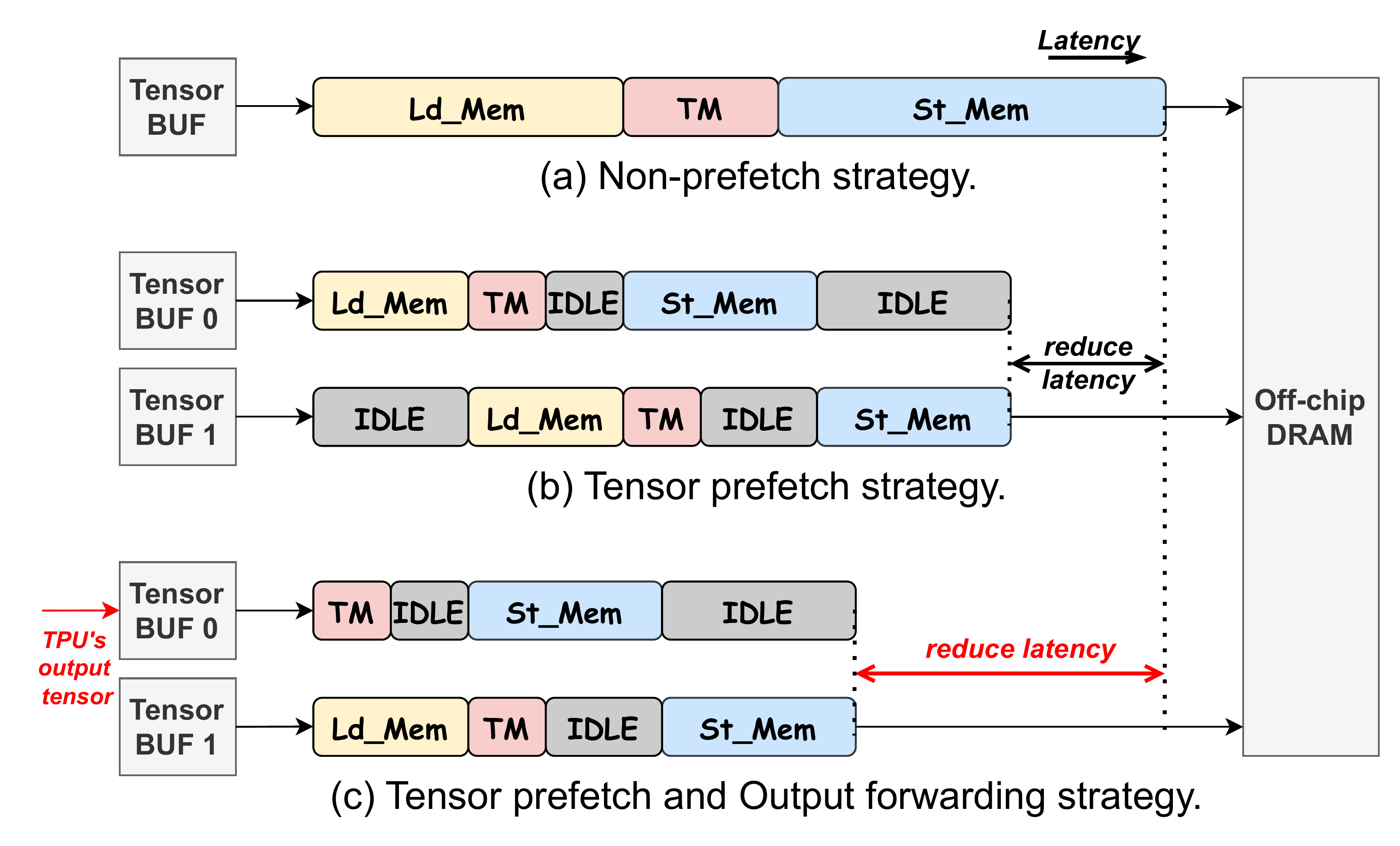}
    \caption{(a) Non-prefetch strategy. (b) Tensor prefetch strategy. (c) Tensor prefetch and Output forwarding strategy.}
    \label{fig:prefetch tensor strategy}
\end{figure}

To maximize the TMU's efficiency in handling these operators, we employ several system-level strategies that enhance both performance and flexibility.

\subsubsection{Tensor Prefetch and Output Forwarding}
As shown in Fig. \ref{fig:prefetch tensor strategy}(b), a tensor prefetching strategy is employed to minimize off-chip memory access latency and enhance TM efficiency. Two on-chip tensor buffers and two TMUs are configured in a double-buffering arrangement, where one buffer processes data while the other concurrently loads or stores datastreams to/from external DRAM. In TMU–TPU cooperative scenarios, a ping-pong mechanism enables the pre-scheduling of partially committed tensors from the TPU, effectively overlapping memory transfers with computation to mask TMU latency.

To further enhance pipeline concurrency, an output forwarding strategy is employed alongside prefetching. As shown in Fig. \ref{fig:prefetch tensor strategy}(c), when the TPU reaches the final stages of a compute-intensive operation (e.g., Conv, as shown in Fig.\ref{fig:architecture}(b)), it begins streaming partial output tensors to the buffer before completing the entire computation. This allows the TMU to initiate subsequent operators—such as PixelShuffle, or Add—early, thereby reducing idle cycles and improving throughput.

\subsubsection{Block-based Manipulation}
Coarse-grained TM operators operate on tensor blocks whose channel dimensions align with the burst width of the AXI interface. The TMU directly manipulates such datastreams and reshapes them within on-chip buffers, thereby increasing data throughput and minimizing memory latency. For fine-grained TM operators, a reconfigurable masking engine (RME) is introduced to optimize data paths and memory utilization, mitigating the performance penalties typically associated with sub-word data manipulation.

\begin{figure*}[!t]
    \centering
    \includegraphics[width=1.0\linewidth]{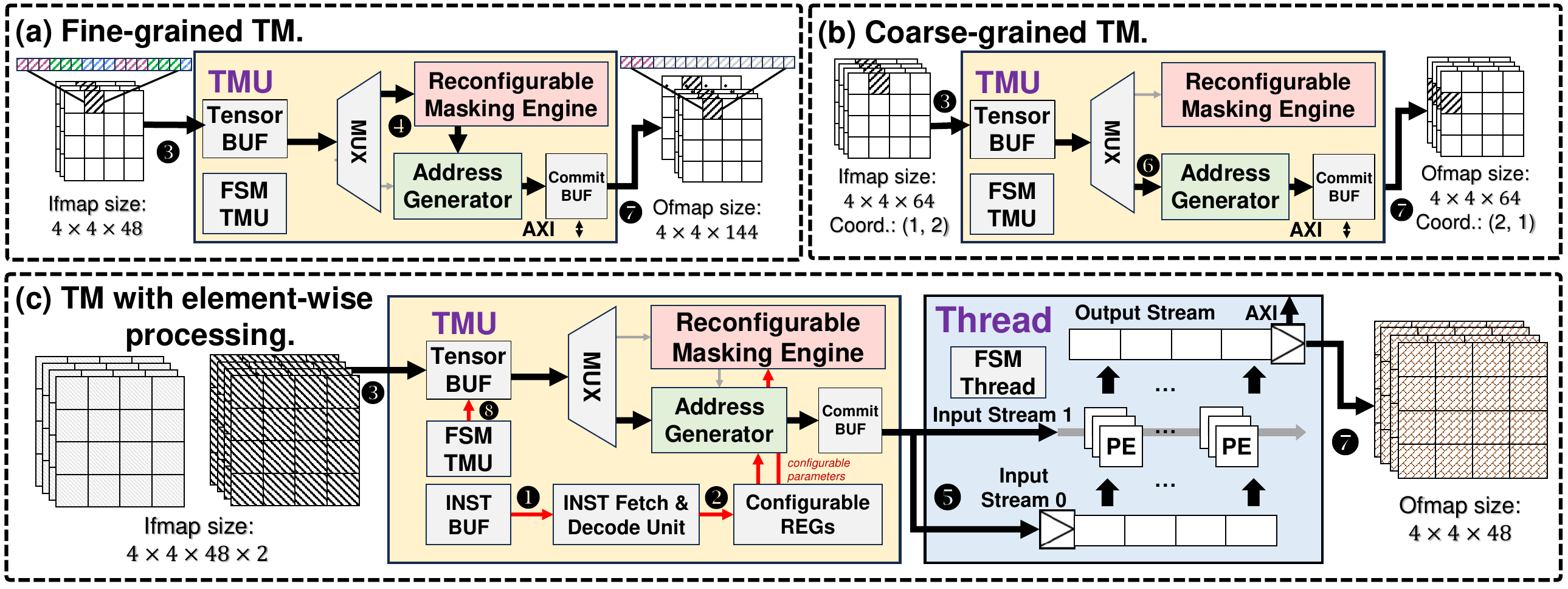}
    \caption{TMU's microarchitecture and dataflows. (a) Fine-grained TM. (b) Coarse-grained TM. (c) TM with element-wise processing.}
    \label{fig:microarchitecture}
\end{figure*}

\begin{figure}[htbp]
    \centering
    \includegraphics[width=1.0\linewidth]{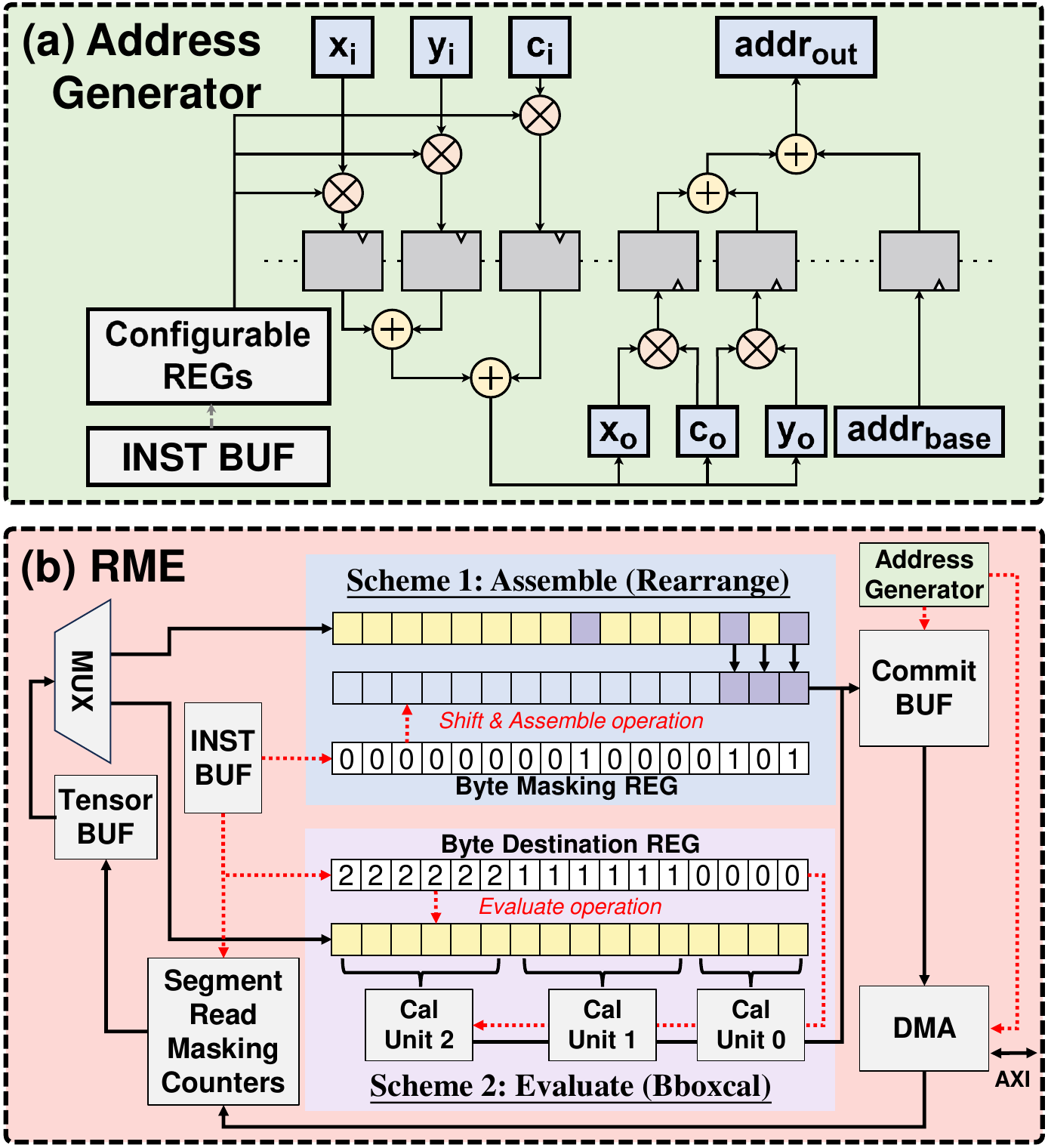}
    \caption{(a) Address generator for coarse-grained TM operators. (b) Reconfigurable masking engine (RME) for fine-grained TM operators.}
    \label{fig:RME}
\end{figure}

\subsection{TMU's Microarchitecture and Dataflows}
As illustrated in Fig.~\ref{fig:microarchitecture}, the TMU integrates the generic execution model, address generator, reconfigurable masking engine (RME), and supporting control modules within a unified architecture. 

A centralized finite-state machine (FSM) orchestrates the execution stages defined in Fig.~\ref{fig:execution_model}, directing the instruction flow and coordinating data movement across the pipeline. To support the heterogeneous characteristics of TM operators, the FSM enables three configurable dataflows corresponding to fine-grained, coarse-grained, and element-wise processing.
\begin{itemize}
    \item \textit{Fetch and Decode:} The TMU retrieves instructions from local memory and decodes them to determine the operator type and operand configuration (see Fig.~\ref{fig:microarchitecture}(c) \ding{182}, \ding{183}).
    \item \textit{Tensor Load:} Input data is fetched via a high-speed bus and stored in on-chip buffers. After potential reshaping or reordering, the datastream is moved to the commit buffer for further processing or storage (see Fig.~\ref{fig:microarchitecture}(a)(b)(c) \ding{184}).
    \item \textit{Fine-grained TM:} This stage performs byte-level manipulation, such as selecting specific bytes (e.g., Rearrange, Bboxcal) or assembling data across segments (e.g., Transpose) to form new datastreams (see Fig.~\ref{fig:microarchitecture}(a) \ding{185}).
    \item \textit{Coarse-grained TM:} This stage handles block-level reshaping operations driven by burst transfers (e.g., 16-byte AXI transactions). The address generator dynamically computes destination addresses to facilitate high-throughput tensor reorganization (see Fig.~\ref{fig:microarchitecture}(b) \ding{187}).
    \item \textit{Element-wise Processing:} This stage supports arithmetic operations such as vectorized Add, Sub, and Mul. The TMU either executes these computations directly or cooperates with processing elements by preprocessing the datastream and forwarding it to global buffers (see Fig.~\ref{fig:microarchitecture}(c) \ding{186}).
    \item \textit{Tensor Store:} The transformed datastream is written back from the commit buffer to off-chip DRAM or shared on-chip memory (see Fig.~\ref{fig:microarchitecture}(a)(b)(c) \ding{188}).
    \item \textit{Branch:} For long tensors spanning multiple iterations, this stage updates memory addresses and buffer offsets to fetch subsequent segments, ensuring uninterrupted execution (see Fig.~\ref{fig:microarchitecture}(c) \ding{189}).
\end{itemize}

\subsubsection{Address Generator} 
As illustrated in Fig. \ref{fig:RME}(a), the address generator implements the matrix-based addressing scheme defined by Eq. \ref{eq:0}, enabling flexible computation of output memory addresses for coarse-grained TM operations. 
At runtime, the TMU instruction stream delivers operator-specific addressing parameters, which are decoded and loaded into dedicated configuration registers. These registers store the elements of matrices $\mathbf{A}$ and $\mathbf{B}$, controlling the affine mapping from input coordinates $(x_i, y_i, c_i)$ to output coordinates $(x_o, y_o, c_o)$.
The address generator executes the transformation in three pipeline stages. First, the row vectors of $\mathbf{A}$ are partitioned into three groups and multiplied with the input index vector. The resulting partial sums are then added to the corresponding entries in $\mathbf{B}$ to form an intermediate vector $\mathbf{C}$. Finally, the output address $addr_{out}$ is computed by combining $\mathbf{C}$ with the base address $addr_{base}$ through post-addition logic.
This output address is used to direct the storage of manipulated tensors to DRAM or to downstream modules. In conjunction with the write stride control, the generator iterates over tensor segments. Internal state registers are updated accordingly to determine whether the current TM instruction has been completed or requires additional iterations.

\subsubsection{Reconfigurable Masking Engine}
To realize fine-grained TM, the RME shown in Fig. \ref{fig:RME}(b) utilizes segment masking counters to acquire valid bus transfers into the tensor buffer.  
Afterwards, It supports two processing approaches for fine-grained TM, namely \textit{assemble} and \textit{evaluate}.

In the assemble scheme, identified bytes in the byte masking register are assembled into a new datastream in the assemble register, which is useful for operators like Rearrange, Rot90, and Transpose. 
The evaluate scheme processes selected bytes to extract result bytes, such as the case of Bboxcal, maximal, or minimal value retrieval from input datastream. The byte destination register maps bytes to specific calculation units, simplifying integration with computational logic. 

While these fine-grained opeators are inherently operation-specific and difficult to build in a generalized structure, the RME mitigates this challenge by offering a structured abstraction for implementation. Specifically, both the \textit{assemble} and \textit{evaluate} schemes adhere to a predefined template comprising three stages: \textit{(i) byte-level masking and indexing}, \textit{(ii) construction and distribution of a new datastream}, and \textit{(iii) conditional routing and commitment under FSM control}. For instance, a new TM op (such as selective value gating) can be mapped to this template by simply configuring its masking rule and output mapping logic, thereby avoiding the need for bespoke dataflow design.

In both schemes, the result datastream is output into the commit buffer under the direction of the address generator. Although both schemes may fill the commit buffer in an interleaved manner, after predictable rounds of processing, a renewed continuous datastream is formed in the commit buffer, which can be streamed to memories through DMA uninterruptedly. The timing of committing is controlled by the FSM tensor store stage.

\begin{table}[!t]
    \centering
    \caption{Operator configuration parameters}
    \label{tab:op config}
    \begin{tabular}{l|c|c|c}
        \hline
        \textbf{TM Op.} & \textbf{Ifmap Size} & \textbf{Ofmap Size} & \textbf{Abbr.} \\
        \hline
        Rearrange & 448$\times$448$\times$3 & 448$\times$448$\times$16 & RR\\
        \hline
        Resize & 448$\times$448$\times$3 & 224$\times$224$\times$3 & RS\\
        \hline
        Bboxcal & 448$\times$448$\times$256 & 3$\times$448$\times$448$\times$85 & BC\\
        \hline
        Transpose & 448$\times$448$\times$64 & 448$\times$448$\times$64 & TS\\
        \hline
        Rot90 & 448$\times$448$\times$64 & 448$\times$448$\times$64 & RT\\
        \hline
        Img2col & 448$\times$448$\times$64 & 446$\times$446$\times$64 & IC\\
        \hline
        PixelShuffle & 448$\times$448$\times$64 & 896$\times$896$\times$16 & PS\\
        \hline
        PixelUnshuffle & 448$\times$448$\times$64 & 224$\times$224$\times$256 & PU\\
        \hline
        Upsample & 448$\times$448$\times$64 & 896$\times$896$\times$64 & US\\
        \hline
        Route & 2$\times$448$\times$448$\times$64 & 448$\times$448$\times$128 & RO\\
        \hline
        Split & 448$\times$448$\times$64 & 2$\times$448$\times$448$\times$32 & SL\\
        \hline
        Add & 448$\times$448$\times$64 & 448$\times$448$\times$64 & AD\\
        \hline
    \end{tabular}
\end{table}

\begin{figure*}[!t]
    \centering
    \includegraphics[width=1.0\linewidth]{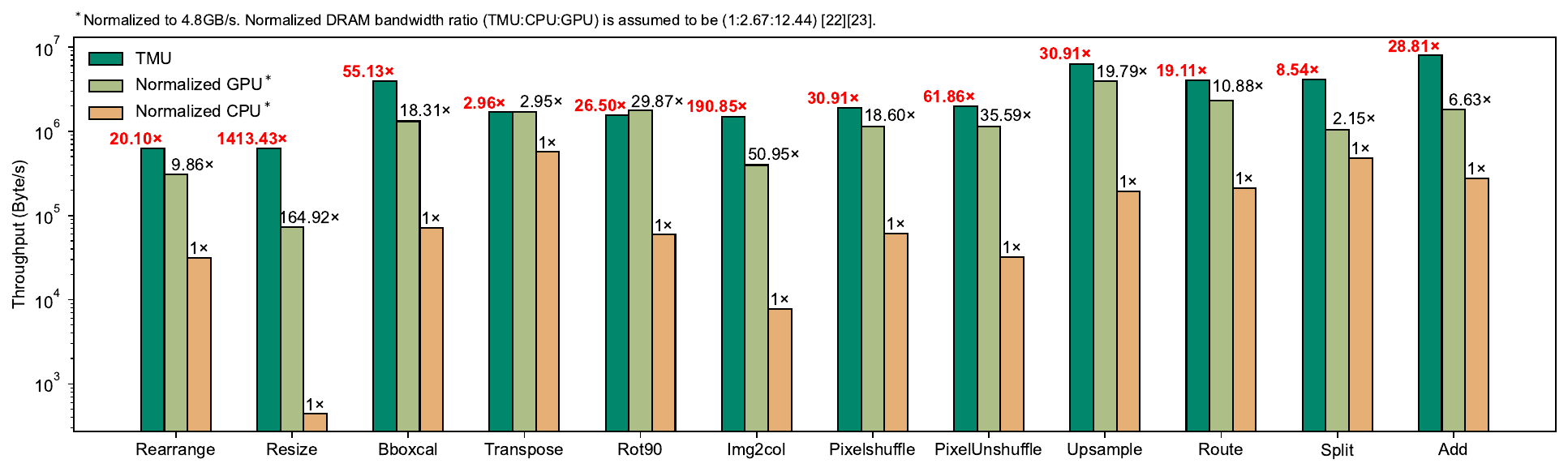}
    \caption{Inference latency benchmarking (TMU Prefetch vs Normalized GPU vs Normalized CPU) for TM operators.}
    \label{fig:operator-level-latency}
\end{figure*}

\section{Experimental Results} \label{sec:experiments}
This section presents our evaluation methodology and summarizes the experimental results of the proposed accelerators in comparison with conventional CPU and GPU platforms.

\begin{table}[!t]
    \centering
    \caption{Model configuration parameters}
    \label{tab:model config}
    \resizebox{\linewidth}{!}{
    \begin{tabular}{c|c|c|c}
        \hline
        \textbf{DNN Type} & \textbf{DNN Model} & \textbf{Ifmap Size} & \textbf{TM Op.} \\
        \hline
        \multirow{5}{*}{CNN} & ESPCN & \multirow{4}{*}{448$\times$448$\times$3} &  RR, PS\\
        \cline{2-2} \cline{4-4}
        & EDSR & & RR, PS, AD\\
        \cline{2-2} \cline{4-4}
        & YOLOv3 & & RR, RO, US, AD, BB\\
        \cline{2-2} \cline{4-4}     
        & YOLOv3-Tiny &  & RR, RO, US, BB\\
        \cline{2-4}
        & YOLOv8 & 640$\times$640$\times$3 & RR, RO, US, AD, SL, BB\\
        \hline
        Transformer & Attention & 64$\times$768 & TS, RO\\
        \hline
    \end{tabular}
    }
\end{table}

\begin{figure}[!b]
    \centering
    \includegraphics[width=0.9\linewidth]{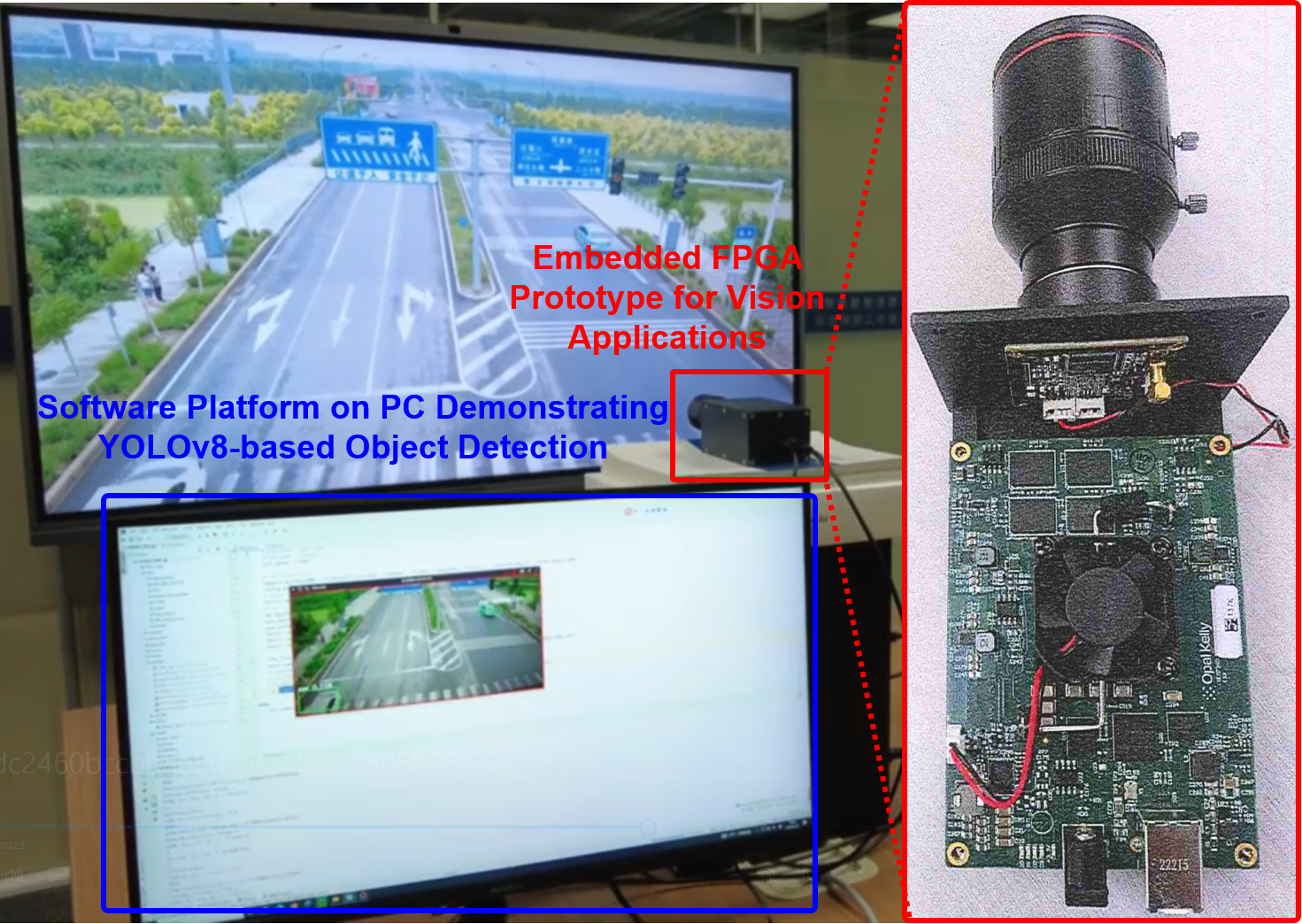}
    \caption{Demonstration of a camera-equipped FPGA hardware prototype and PC-based software platform for YOLOv8 object detection}
    \label{fig:FPGA}
\end{figure}

\subsection{Experimental Setup}
\subsubsection{Implementation}
The proposed TMU and its coupled TPU are fully implemented in Verilog, verified through VCS simulation, and integrated into an in-house SoC infrastructure. The complete system is deployed on a Xilinx Kintex-7 410T FPGA board, which communicates with the host environment via python-based interfaces and a USB 3.0 data link. Logic synthesis is performed using Synopsys Design Compiler with SMIC $40~\mathrm{nm}$ low leakage standard cell libraries, targeting a clock frequency of $300~\mathrm{MHz}$.
Fig. \ref{fig:FPGA} shows the physical hardware prototype, featuring a camera-equipped FPGA board running real-time object detection with YOLOv8\cite{varghese2024yolov8}, along with a PC-based software interface for system interaction and visualization.

\subsubsection{Software Platform Setup}
The evaluated neural networks are processed using an in-house AI compilation toolchain built with Python 3.6, TensorFlow 2.6, and NumPy 2.2. This toolchain parses each model to extract its computational graph and operator dependencies, applies post-training quantization (PTQ) using TensorFlow Lite to convert weights to INT8 precision, and transforms the result into an intermediate representation (IR) for hardware-specific optimization. The optimized models are then deployed to the target platforms. Using this toolchain, we conduct a comprehensive inference latency evaluation at both the TM operator and application levels across TPU, TMU, Jetson TX2, and Raspberry Pi 4. 

\begin{figure*}[!ht]
\centering
    \subfigure[Inference latency benchmark for the entire neural network.]{ 
    \begin{minipage}[t]{0.32\textwidth}
    \centering
    \includegraphics[height=8cm]{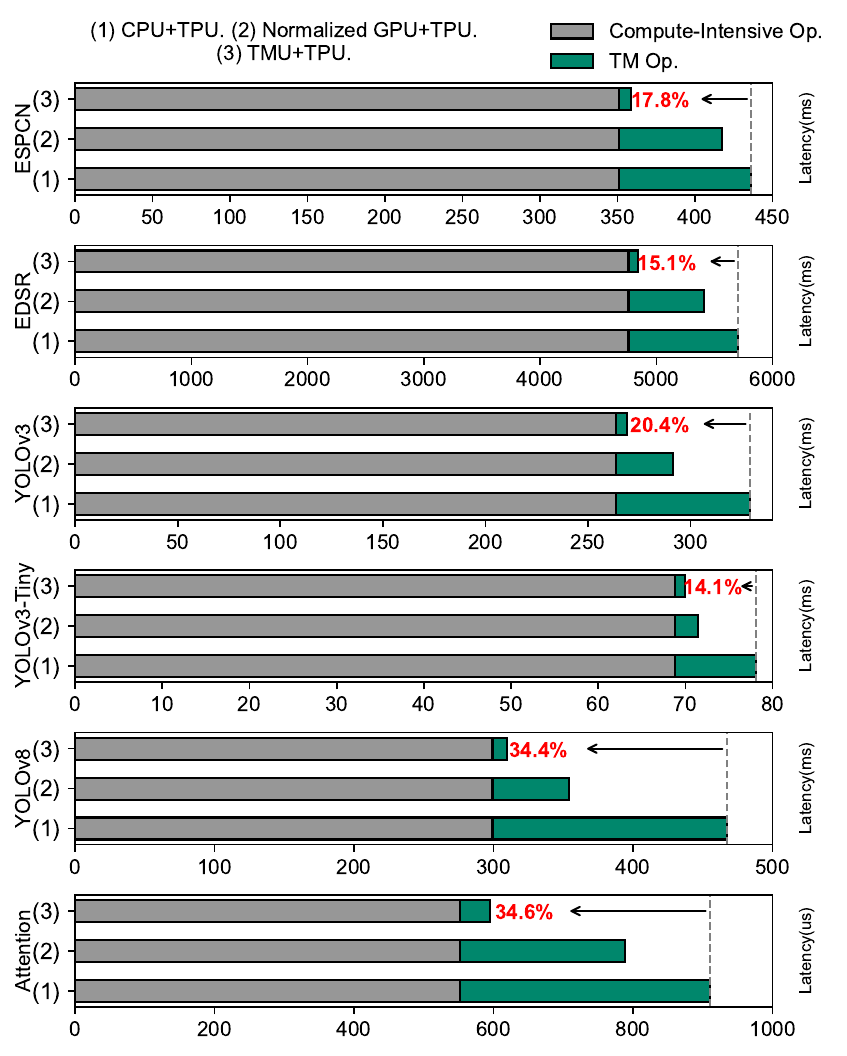}
    \end{minipage}
    }
    \subfigure[Inference latency benchmark for TM operators only.]{ 
    \begin{minipage}[t]{0.64\textwidth}
    \centering 
    \includegraphics[height=8cm]{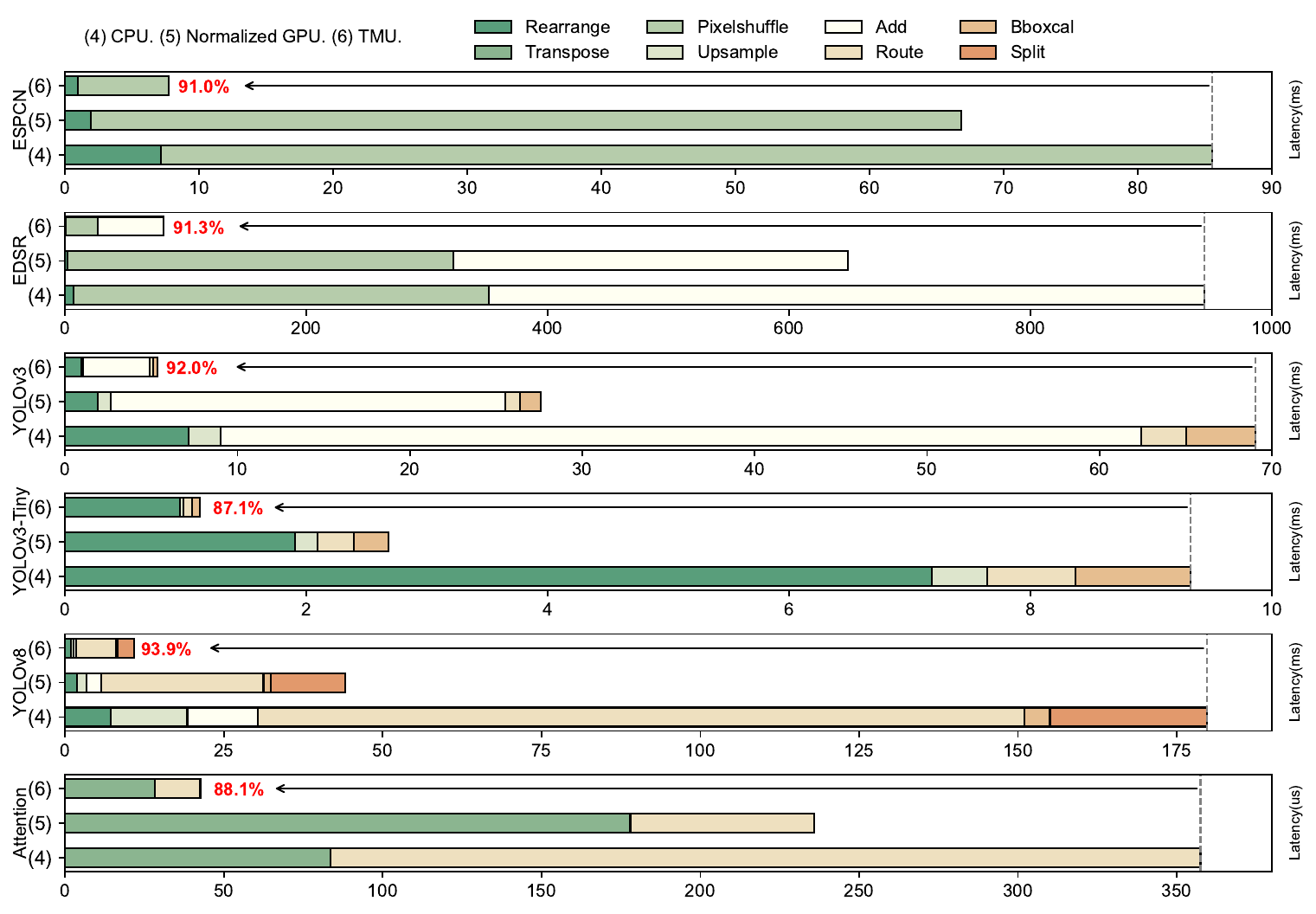}
    \end{minipage}
    }
    \caption{Graphical representation of typical TM operators adopted in state-of-the-art neural networks.}   
    \label{fig:application-level-latency} 
\end{figure*}

\subsection{Performance Evaluation}\label{AA}
\subsubsection{TM operators benchmarking}
We evaluated the performance of various TM operators on the TMU, comparing it to a $1.3 \mathrm{GHz}$ NVIDIA Pascal GPU (Jetson TX2) and a $1.5 \mathrm{GHz}$ ARM Cortex-A72 CPU (Raspberry Pi 4 Model B) using official TensorFlow library functions. 
The configuration parameters of TM operators are listed in Table \ref{tab:op config}.
The Jetson TX2 and Raspberry Pi 4 Model B were chosen for their relevance in edge computing, providing high programmability for executing TM operators, unlike other typical accelerators, which often lack support for a wide range of TM operators, limiting their applicability in such tasks. Additionally, both devices employ a complex multi-level caching mechanism for data movement, which contrasts with the near-memory DMI approach adopted by the TMU.

Note that the performance of TM operators is heavily constrained by DRAM bandwidth. To enable fair comparison across platforms, the measured performance of the CPU (with a DRAM bandwidth of $12.8 \mathrm{GB/s}$\cite{8188629}) and the GPU (with a bandwidth of $59.7 \mathrm{GB/s}$\cite{nvidiaTX2Datasheet}) is normalized to match the DRAM bandwidth of the TMU, which is $4.8 \mathrm{GB/s}$. This normalization ensures that observed performance differences reflect architectural design efficiency rather than bandwidth disparities, enabling a more meaningful and bandwidth-fair comparison across platforms.

As shown in Fig. \ref{fig:operator-level-latency}, the TMU demonstrates substantial throughput improvements,  
It achieves up to $1413.43\times$ in Resize and $61.86\times$ in PixelUnshuffle than CPU, consistently outperforming the normalized GPU and CPU. The TMU excels in various fine-grained TM operators, such as $55.13\times$ in Bboxcal. Significant gains are also observed in element-wise operators like $19.11\times$ in Route and $28.81\times$ in Add. These results highlight the TMU's potential in performance-boosting of TM operators.
The only TM operator where the TMU underperforms GPU is Rot90, due to the time-intensive data disassembling and reassembling, particularly between the width and channel dimensions, which can be further optimized in the ongoing TMU implementation.

\subsubsection{Application-level benchmarking}
To ensure fairness and reproducibility in application-level benchmarking, we adopt a unified evaluation methodology. On CPU and GPU platforms, TensorFlow implementations are instrumented with custom profiler context blocks to measure the execution time of TM layers. All reported results exclude cross-platform communication overhead to isolate computation performance.
In this setup, compute-intensive operators, including convolutions and associated transformations (e.g., \textit{Img2col}), are executed on the TPU, while TM operators are handled by the TMU, CPU, and GPU platforms. To highlight the TMU’s contribution to neural network acceleration, the impact of \textit{Img2col} is excluded from the application-level benchmarking.

Execution times for six typical deep learning networks featuring various TM operators are shown in Fig. \ref{fig:application-level-latency}(a). These applications include ESPCN\cite{talab2019super}, EDSR\cite{lim2017enhanced}, YOLOv3\cite{redmon2018yolov3}, YOLOv3-Tiny\cite{adarsh2020yolo}, YOLOv8\cite{varghese2024yolov8}, and Attention\cite{vaswani2017attention}.
The configuration parameters of application-level benchmarking are listed in Table \ref{tab:model config}.
Combined with our in-house TPU \cite{li2024low}, the TMU outperformed a conventional CPU (even without performance normalization) in inference latency benchmark for the entire neural network, achieving speedup of $17.8\%$, $15.1\%$, $20.4\%$, $14.1\%$, $34.4\%$, and $34.6\%$ for ESPCN, EDSR, YOLOv3, YOLOv3-Tiny, YOLOv8, and Attention, respectively. 
Fig. \ref{fig:application-level-latency}(b) shows the accumulated latency for all TM operators, where the TMU demonstrated significant reductions in latency, achieving reductions of $91.0\%$, $91.3\%$, $92.0\%$, $87.1\%$, $93.9\%$, and $88.1\%$ for ESPCN, EDSR, YOLOv3, YOLOv3-Tiny, YOLOv8, and Attention, respectively. 

\subsection{Physical Overheads}
Table \ref{tab:physical} provides a comparison between the proposed TMU and several state-of-the-art DMI accelerators. It is important to note that the reported physical parameters do not include the DRAM controller or interface circuitry. 
Compared to prior works, this design is the first DMI accelerator to serve as a generic design template with reconfigurable registers capable of supporting a wide range of TM operators. 
In addition, the proposed TMU demonstrates clear advantages in both area and power efficiency. After normalization to a $40~\mathrm{nm}$ process, it achieves the smallest area footprint ($0.019~\mathrm{mm}^2$), which is $15.3\times$ smaller than that of AME\cite{sugawara2022data}. While the power comparison is based on reported values without frequency normalization, the TMU still consumes $1.52\times$ less power ($2.7~\mathrm{mW}$ vs $4.1~\mathrm{mW}$).
Note that the adopted normalized area and ratio refers to paper \cite{mo202112} and report \cite{TSMC2019}.

Furthermore, this work supports a broader range of functions (\ding{172}-\ding{179}), while AME \cite{sugawara2022data} and ECNN \cite{huang2019ecnn} lacks functional support. These advantages underscore the efficiency of the proposed architecture in minimizing power consumption and area footprint while maintaining competitive performance and extensive functional support.
Additionally, The TMU+TPU combination achieves high-performance improvement and a series of TM operators with only $0.07\%$ extra overhead of the TPU's area, this indicates that the TMU contributes significantly to system-level optimization.

\begin{table}[!ht]
\begin{center}
\caption{Comparison with state-of-the-art data-move-intensive accelerators.} \label{tab:physical}
    \centering
    \begin{threeparttable}
    \begin{tabular}{|>{\columncolor{gray!20}}c|c|c|c|}
    \hline  
    \rowcolor{gray!20}
    Accelerators & ECNN\cite{huang2019ecnn}\dag & AME\cite{sugawara2022data} & This Work\\
    \hline
    Technology & $40~\mathrm{nm}$ & $7~\mathrm{nm}$ & $40~\mathrm{nm}$\\
    \hline
    Frequency($\mathrm{MHz}$) & 250 & 2100   & 300\\
    \hline
    Area($\mathrm{mm^2}$) & 2.26 & 0.034 & 0.019\\
    \hline
    Power($\mathrm{mW}$) & 100 & 4.1 & 2.7\\
    \hline
    Normal. Area\ddag($\mathrm{mm^2}$) & 2.26 & 0.291 & 0.019\\
    \hline
    Reconfigurability & \ding{56} & \ding{56} & \ding{52}\\
    \hline
    Function\S & \ding{175},\ding{177},\ding{180} & N/A & \ding{172}-\ding{179}\\
    \hline
    \end{tabular}
    \end{threeparttable}
    \\[\baselineskip] 
    
    \resizebox{\linewidth}{!}{  
    \begin{tabular}{|>{\columncolor{gray!20}}c|c|c|c|c|c|}
    \hline
    \rowcolor{gray!20}
    &Freq.& Area & Power & DRAM & \#.MACs\\
    \hline
    Integrated & 300& 26.96$~\mathrm{mm^2}$&1.83$~\mathrm{W}$&1200$~\mathrm{MT/s}$&4096\\
    TPU \cite{li2024low}&$\mathrm{MHz}$& post P\&R & w. DRAM & DDR3 &(int8)\\
    \hline
    \end{tabular}
    }

    \begin{threeparttable}
    \begin{tablenotes}    
        \footnotesize
        \item\dag ECNN\cite{huang2019ecnn} includes Src/SrcS/Dst Reorder, ADDE/ACCI, MUXA/MUXC, Block Buffer File, and other modules.
        \item\ddag Normalized area ratio ($7~\mathrm{nm}$:$40~\mathrm{nm}$) = (1:8.57) \cite{mo202112}, \cite{TSMC2019}. Normalized to $40~\mathrm{nm}$, $300~\mathrm{MHz}$.
        \item\S \ding{172}: Rearrange, \ding{173}: Resize, \ding{174}: Bboxcal, \ding{175}: Rot90 and Transpose, \ding{176}: Img2col, \ding{177}: PixelShuffle and PixelUnshuffle, \ding{178}: Upsample, \ding{179}: Route, Split, and Add, \ding{180}: Downsample.
      \end{tablenotes} 
    \end{threeparttable}
\end{center}
\end{table}

\section{Conclusion}
\label{sec:conclude}
This work presents TMU, a reconfigurable near-memory TMU designed to accelerate DMI operators in modern AI workloads. 
By addressing the frequently neglected latency bottlenecks introduced by TM operators, TMU significantly enhances system throughput across multiple state-of-the-art models.
Despite its compact hardware footprint, the TMU delivers substantial performance gains and demonstrates strong scalability and integration potential within high-throughput AI SoCs.

\renewcommand{\baselinestretch}{1}

\end{document}